\documentclass[a4paper,11pt]{article}
\pdfoutput=1 

\usepackage{jheppub} 
\usepackage{mathrsfs}
\graphicspath{{Figures/}}
\newcommand{\beq}{\begin{equation}}
\newcommand{\eeq}{\end{equation}}

\let\Om=\Omega

\newcommand{\vc}{volume-complexity}
\newcommand{\ep}{effective potential}

\title{Generalized Volume Complexity in Gauss-Bonnet Gravity: Constraints and Phase Transitions}
\author[a,*]{Xuanhua Wang,} 
\author[b,*]{Ran Li,}
\author[c,d,*]{Jin Wang \note[*]{Corresponding authors}}
\affiliation[a]{Center for Theoretical Interdisciplinary Sciences, Wenzhou Institute, University of Chinese Academy of Sciences, Wenzhou, Zhejiang 325001, China}
\affiliation[b]{Department of Physics, Qufu Normal University, Qufu, Shandong 273165, China}
\affiliation[c]{Department of Chemistry, Stony Brook University, Stony Brook, NY 11794, USA}
\affiliation[d]{Department of Physics and Astronomy, Stony Brook University, Stony Brook, NY 11794, USA}

\emailAdd{wangxh@ucas.ac.cn}
\emailAdd{liran@qfnu.edu.cn}
\emailAdd{jin.wang.1@stonybrook.edu}

\abstract{It has been proposed that quantum complexity is dual to the volume of the extremal surface, the action of the Wheeler-DeWitt patch, and the spacetime volume of the patch. Recently, a generalized volume-complexity observable was formulated as an equivalently good candidate for the dual holographic complexity. This proposal is abbreviated as ``complexity=anything." This proposal offers greater flexibility in selecting extremal surfaces and evaluating physical quantities, e.g., volume or action, on these surfaces. In this study, we explore the 'complexity=anything' proposal for Gauss-Bonnet black holes in asymptotic anti-de Sitter space in various dimensions. We demonstrate that this proposal guarantees the linear growth of the generalized volume at late times, regardless of the coupling parameters for four-dimensional Gauss-Bonnet gravity. However, this universality does not hold for higher dimensions. Moreover, discontinuous deformations of the extremal surfaces emerge when multiple peaks exist in the effective potential, which is reminiscent of a phase transition. Additionally, we present constraints on the coupling parameters of five-dimensional models in order for the generalized volume to be a viable candidate for holographic complexity.}

\begin{document} 
\maketitle

\section{Introduction}
The holography of black holes has triggered much excitement and progress in our understanding of black holes over the last decade. The core of this treatment is the validity of the central dogma that black holes exhibit unitary evolutions \cite{Almheiri:2020cfm}. It has been demonstrated in both replica wormhole calculations as well as qubit models that the entropy of a black hole is encoded in the area of its horizon after the Page time \cite{Page:1993df,Page:1993wv}. The nature of the Page transition has been hypothesized, for example, as the emergence of Euclidean wormholes \cite{penington2022replica,almheiri2019entropy}, the quantum teleportation induced by effective measurements \cite{wang2023entanglement}, and recently, the transition of the information channels \cite{li2023high,li2023retrieving}. While the entropy of a black hole increases only for a finite duration before reaching thermalization, the volume of its interior and the length of the wormhole for a two-sided AdS black hole continue to grow \cite{Hartman:2013qma}. Consequently, it was suggested that a new quantum information measure is needed to capture the feature. Various concepts have emerged in this endeavor, such as linking volumes to this complexities of decoding information within the black hole, and several explicit calculations have been developed to demonstrate this duality \cite{Susskind:2014rva,Stanford:2014jda,Couch:2016exn,Belin:2021bga,Belin:2022xmt,Wang:2023eep,Yang:2016awy,An:2018dbz,Ghodsi:2020qqb,Mandal:2022ztj,Babaei-Aghbolagh:2021ast,Cai:2020wpc,HosseiniMansoori:2017tsm,Auzzi:2023qbm,Emparan:2021hyr,Zolfi:2023bdp,Fan:2019mbp}. 

Quantum complexity is a measure of least steps to complete a task by applying gates only from a set of finite simple operations. The ambiguity in determining the exact value of complexity arises from the arbitrariness of the choice of gate set and the cost assigned to each gate. It was argued that, instead of a shortcoming, this ambiguity connects nicely with the ambiguity of rescaling freedom of the length/volume in its holographic dual \cite{Belin:2021bga,Belin:2022xmt}. The scaling of the complexity with the size of the problem is a robust property, and the quantum circuit complexity is shown to grow linearly with the number of Haar-random two-qubit quantum gates \cite{haferkamp2022linear}.

Inspired by the ER$=$EPR proposal \cite{Maldacena:2013xja}, the observation of linear growth of the wormhole length and the switchback effect of external perturbations, the quantum complexity is conjectured to be dual to volume of the codimension-one maximal slice (Complexity-Volume proposal, or CV) \cite{Susskind:2014rva,Stanford:2014jda}, gravitational action in the Wheeler-DeWitt patch (Complexity-Action proposal, or CA) \cite{Brown:2015lvg,Brown:2015bva,Carmi:2016wjl} and the spacetime volume of the Wheeler-DeWitt patch (Complexity-Volume proposal 2.0, or CV2.0) \cite{Couch:2016exn}. In the CV conjecture, the holographic complexity is conjectured to be dual to the maximal volume of the hypersurface anchored at the boundary CFT slice $\Sigma_{\mathrm{CFT}}$, viz.
\begin{gather}
    C_{V}(\Sigma_{\mathrm{CFT}})=\max\limits_{\partial\mathcal{B}=\Sigma_{\mathrm{CFT}}}\left[\frac{\mathcal{V}(\mathcal{B})}{G_N \ell_{\mathrm{bulk}}}\right]\,,
\end{gather}
where $\mathcal{B}$ is the bulk hypersurface. The CA conjecture offers an alternative to the CV proposal and the states that the complexity is represented by the integral of the gravitational action in the Wheeler-DeWitt patch, which is defined as the domain of dependence of the slice $\mathcal{B}$. It is expressed as:
\begin{gather}
     C_{A}(\Sigma_{\mathrm{CFT}})=\frac{I_{\mathrm{WDW}}}{\pi \hbar}\,.
\end{gather}
The CA proposal was investigated in various models. See Refs.~\cite{,Mounim:2021ykr,Sajadi:2022uwg,Anegawa:2023wrk,Meng:2018vtl,Omidi:2020oit,Akhavan:2019zax,Bravo-Gaete:2020lzs,Alishahiha:2018lfv,Yekta:2020wup} for a non-exhaustive list of works. The CV2.0 proposal combines the CV and CA conjectures and proposes that the complexity is simply given by the spacetime volume of the Wheeler-DeWitt patch anchored at the given boundary states,
\begin{gather}
    C_{V2}(\Sigma_{\mathrm{CFT}})=\frac{V_{\mathrm{WDW}}}{G_N\ell^2_{\mathrm{bulk}}}\,.
\end{gather}
Recently, it was realized that the dual of holographic complexity, which exhibits linear growth and the switchback effect at late times, can be extended to an infinite family of observables of either codimension zero or one \cite{Belin:2021bga,Belin:2022xmt,Jorstad:2023kmq}. For codimension-one observables, which is the focus of this paper, the generalized volume is given by
\begin{gather}
    C_{gen}(\tau)=\max_{\partial\Sigma(\tau)=\Sigma_{CFT}}\left(\frac{1}{G_N L}\int_{\Sigma(F_2)}d^{D-1}\sigma\sqrt{h}F_1(g_{\mu\nu};X^\mu(\sigma))\right)\,,
    \label{Cgen}
\end{gather}
where the slice $\Sigma(F_2)$ is the extremal surface for a scalar function $F_2$ and $\tau$ is the boundary time. Notably, $F_2$ can be any scalar function, and it is not required to be the same as $F_1$ in the integrand. It is shown that at the late times, different choices of the functions $F_2$ result in the same late-time behavior for the generalized volume \cite{Belin:2021bga}. Therefore, demonstrating the simpler case $F_1=F_2$ should suffice for the discussion. The proposal of Complexity=Anything (CAny) is useful for probing the singularities of black holes by tuning the coupling parameters \cite{Belin:2022xmt,Jorstad:2023kmq}. Recent development extends the CAny to de-Sitter spaces \cite{Aguilar-Gutierrez:2023zqm}, charged black holes \cite{Wang:2023eep,Jiang:2023jti}, and two-sided Hyperscaling Violating black branes \cite{Omidi:2022whq}. The details of this proposal will be delineated in the subsequent sections. 

The goal of this paper is to explore the generalized CAny proposal of the codimension-one surfaces for Gauss-Bonnet black holes in four dimensions and higher. Importantly, the original CAny proposal was formulated in the Einstein gravity. Generally, its equivalence in the Gauss-Bonnet gravity may have contributions from more sources such as the extrinsic curvatures in the CV proposal \cite{Hernandez:2020nem} or the Gauss-Bonnet coupling $\alpha$. The quest for all possible generalizations is beyond the scope of the current paper and requires further investigations. In this study, we show that the CAny proposal can be directly generalized to the Gauss-Bonnet gravity. Specifically, we show that in four dimensions, the effective potentials for the Gauss-Bonnet black holes, which dictate the time evolution of the extremal surfaces, are qualitatively different from those in higher dimensions due to the existence of multiple horizons. This feature in 4D Gauss-Bonnet black holes guarantees the linear growth of the generalized volume at late times, regardless of the chosen coupling parameters. This trait of universal linear growth is absent for Gauss-Bonnet black holes in dimensions higher than four. In addition, we explore the scenario when multiple extremal surfaces exist and show that different rates of volume growth can be assigned to the surfaces. We demonstrate the time evolution of conserved momenta corresponding to different configurations of the extremal surfaces and illustrate the conditions for the generalized volume to be qualified as the dual of the holographic complexity. In Sec.~\ref{sec:gb}, we provide background information on the essential features of Gauss-Bonnet black holes. In Sec.~\ref{sec:cany}, we review the CAny proposal and illustrate its application to codimension-one observables. In Sec.~\ref{sec:4d}, we concentrate on the CAny proposal in 4D Gauss-Bonnet gravity. The analysis of 5D Gauss-Bonnet gravity is provided in Sec.~\ref{sec:5d}. We summarize our results in the last section.

\section{Gauss-Bonnet-AdS black holes}\label{sec:gb}
The Gauss-Bonnet Lagrangian is a natural generalization of general relativity which modifies the equations of motion in dimensions larger than four. In 4D, the Gauss-Bonnet term is topological and does not contribute to the equations. Several techniques have been developed to establish a consistent nontrivial 4D Gauss-Bonnet gravity, and the related issues have triggered vigorous discussions \cite{fernandes20224d}. Though issues in 4D are more subtle and not entirely understood, it is still worthwhile to explore the full consequences of the theory.

The 4D Gauss-Bonnet-AdS black holes have two horizons, which implies that the central singularity is time-like \cite{glavan2020einstein,Fernandes:2020nbq,Dadhich:2020ukj}. In spacetime dimensions $D\ge 5$, the solution has only one horizon, similar to the Schwarzschild solution. The topology of Gauss-Bonnet black holes is an active research topic and has been investigated in Refs.~\cite{Wei:2022dzw,Liu:2022aqt,Li:2023ppc,Li:2023men,Li:2020spm}. The action for the GB-AdS black hole is given by \cite{Boulware:1985wk}
\begin{gather}
    \mathcal{S}=\frac{1}{16\pi}\int d^D x \sqrt{-g} (R-2\Lambda +\tilde\alpha \mathscr{L}_{GB})+S_{GBM}\,,
\end{gather}
where $\mathscr{L}_{GB}=R^2-4R_{\mu\nu}R^{\mu\nu}+R_{\mu\nu\rho\sigma}R^{\mu\nu\rho\sigma}$ is the Gauss-Bonnet term, $\Lambda=-\frac{(D-1)(D-2)}{2L^2}$ and $S_{GBM}$ is the 
Gauss-Bonnet-Myers term, which is the generalization of Gibbons-Hawking-York boundary term \cite{Myers:1987yn}. For simplicity, we donote $\alpha=\tilde\alpha (D-3)(D-4)$. Note that the GB coupling $\tilde\alpha$ is scaled as $\frac{1}{D-4}$ which cancels out the $(D-4)$ factor in the definition of $\alpha$. Taking the limit as $D\rightarrow 4$ yields the effective equation of the Einstein-Gauss-Bonnet-AdS model in $4$D. The causality bound from the boundary CFT requires $|\alpha/L^2|\ll 1$ \cite{buchel2009causality}. On the other hand, a well-defined vacuum of the theory also requires that $0\le \frac{4\alpha}{L^2}\le 1$.  For spherically symmetric solution, the metric is given by
\begin{gather}
    ds^2=-f(r)dt^2+f(r)^{-1}dr^2+r^2 d\Omega_{D-2}^2\;,\\
\end{gather}
where $\Omega_{D-2}$ is the phase-space volume of the $(D-2)$-dimensional sphere. In Eddington-Finkelstein coordinates, it reads
\begin{gather}
     ds^2=-f(r)dv^2+2\,dv\,dr+r^2 d\Omega_{D-2}^2\,,
\end{gather}
where the infalling coordinate $v=t+r_*(r)$ with $r_*(r)=-\int_{r}^{\infty} \frac{dr'}{f(r')}$, and the lapse function $f(r)$ is given by
\begin{gather}
    f(r)=1+\frac{r^2}{2\alpha}\left[1-\sqrt{1+4\alpha\left(\frac{w}{r^{D-1}}
-\frac{1}{L^2}\right)}\right]\;.
\end{gather}
Here, $ w= 16\pi GM/\left((D-2)\Om_{D-2}\right)$ is the re-scaled mass of the black hole. The location of the horizon satisfies $f(r_h)=0$, viz.,
\begin{gather}
    w=r_h^{D-3}(1+\frac{r_h^2}{L^2}+\frac{\alpha}{r_h^2})\,.
\end{gather}
The temperature of the black hole is given by
\begin{gather}
    T_{BH}=\left.\frac{f(r)'}{4\pi}\right|_{r=r_h}=\frac{(D-3)r_h^2+(D-1)r_h^4/L^2+(D-5)\alpha}{4\pi r_h \left(r_h^2+2\alpha\right)}\,.
\end{gather}

\section{CAny proposal}\label{sec:cany}

For the codimension-one observable, CAny generalizes the CV and the CA proposals to observables of the form
\begin{gather}
    C_{gen}(\tau)=\max_{\partial\Sigma(\tau)=\Sigma_{CFT}}\left(\frac{1}{G_N L}\int_{\Sigma}d^{D-1}\sigma\sqrt{h}F_1(g_{\mu\nu};X^\mu(\sigma))\right)\,,
    \label{Cgen}
\end{gather}
where the slice $\Sigma$ is the extremal surface for the scalar function $F_2$ which is not necessarily the same as the scalar function $F_1$ in the integrand and $\tau$ is the boundary CFT time with $\tau=2 t_L=2t_R$. The codimension-one extremal slice can be readily extended to the codimension-zero case, where the integration spans the spacetime region enclosed by two extremal slices, along with the contributions from the two codimension-one slices. These observables exhibit linear growth with time at late times and display the switchback effect.  In this study, we investigate the CAny conjecture in Gauss-Bonnet gravity using the square of Weyl tensors and discuss the conditions under which this conjecture holds.

\begin{figure}
    \centering
    \includegraphics[width=0.6\textwidth]{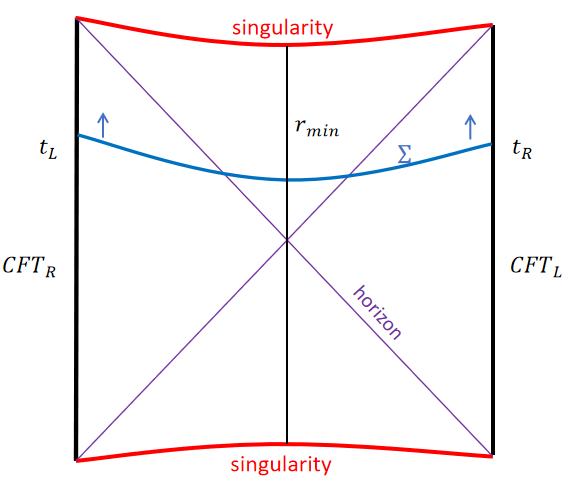}
    \caption{Extremal slices}
    \label{fig:1}
\end{figure}

For simplicity, we pick $F_1=F_2$ and consider the spherically symmetric scalar functions. Choosing the parameterization $(v(\sigma),r(\sigma),\Vec{\Omega})$, we have
\begin{gather}
    C_{gen}(\tau)=\frac{\Omega_{D-2}}{G_N L}\int_{\Sigma} d\sigma \sqrt{-f(r)\dot{v}^2+2\dot{v}\dot{r}}\, F_1(r)\,.
\end{gather}
Note that by choosing $F_1=1$, it returns to the CV proposal. One can identify the generalized complexity as the action and the integrand as the Lagrangian. From the classical equation of motion $\left.\frac{\partial S}{\partial q^i}=\frac{\partial L}{\partial \dot q^i}\right|_{\mathrm{\partial \Sigma}}$, we obtain 
\begin{gather}
    \frac{dC_{gen}}{d\tau}=\left.\frac{1}{2}\frac{\partial (\sqrt{h}F_1)}{\partial \dot{t}}\right|_{\partial \Sigma(\tau)}=\left.\frac{1}{2}P_t \right|_{\partial \Sigma(\tau)}  \,.
\end{gather}
The generalized complexity is diffeomorphism invariant and we employ our freedom of reparametrization to pick the gauge condition
\begin{gather}
    \sqrt{-f(r)\dot{v}^2+2\dot{v}\dot{r}}=F_1(r) \left(\frac{r}{L}\right)^{D-2}\,.
\end{gather}
The momentum $P_\nu$ conjugate to the infalling time $\nu$ is
\begin{gather}
    P_\nu=\frac{\partial \mathcal{L}}{\partial \dot\nu}=\dot{r} -f(r) \dot{\nu}\,,
\end{gather}
where the dots are the derivatives taken with respect to the parameter $\sigma$ which increases from the left AdS boundary to the right AdS boundary.
The above two equations give the extremality conditions:
\begin{align}
    & \dot{r} =\pm \sqrt{P_\nu^2+f(r)F_1^2\left(\frac{r}{L}\right)^{2(D-2)}}\,,\\
    & \dot\nu=\frac{1}{f(r)}\left(-P_\nu \pm \sqrt{P_\nu^2++f(r)F_1^2\left(\frac{r}{L}\right)^{2(D-2)}}\right)=\frac{1}{f(r)}(-P_\nu +\dot r)\,.
    \label{Eq:randv}
\end{align}
Comparing the equations with the motion of a classical particle, we obtain the equations of motion
\begin{gather}
    {\dot{r}}^2+\tilde{U}(r)=P_{\nu}^2 \quad \mathrm{with} \quad \tilde{U}(r)=-f(r)F_1(r)^2\left(\frac{r}{L}\right)^{2(D-2)}\,.
\end{gather}
Note that 
\begin{align}
    \dot\nu=\dot t+\frac{dr_*}{dr}\frac{dr}{d\sigma}=\dot t+\frac{\dot r}{f(r)}\,,
\end{align}
and the conjugate momentum $P_\nu$ is conserved for a specific boundary time $\tau$. 
Combining it with the above extremality condition Eq.~\eqref{Eq:randv}, we have 
\begin{align}
    \dot t=-\frac{P_\nu}{f(r)}\,.
\end{align}
From the equations of $\dot t$ and $\tilde U(r)$, we can recast the generalized volume as
\begin{gather}
    C_{gen}(\tau)=\frac{\Omega_{D-2}}{G_N L}\int_\Sigma^\tau dt\ \frac{\tilde U(r(t))}{P_\nu}\,.
\end{gather}
Here the integration is taken on the extremal surface so that $P_\nu$ in the denominator is a constant. The boundary time $\tau$ is related to the conjugate momentum by
\begin{gather}
    t=\tau/2=-\int_{r_{\mathrm{min}}}^\infty dr \frac{P_\nu}{f(r)\sqrt{P_\nu^2-\tilde{U}(r)}}\,,
\end{gather}
where $r_{\mathrm{min}}$ is the minimal radius lying on the timelike surface $t=0$. The growth rate of the complexity
\begin{gather}
    \frac{d C_{gen}}{d\tau}=\frac{\Omega_{D-2}}{G_N L}P_{\nu}(\tau)\,
\end{gather}
can be calculated from the effective potential. The linear growth of $C_{gen}$ at late times is dedicated by the condition that $\lim_{\tau\rightarrow \infty} P_\nu(\tau):=P_\infty$ is constant and the effective potential $\tilde U(r)$ has a local maximum inside the horizon. In this case,
\begin{gather}
    \lim_{\tau\rightarrow\infty} \frac{d C_{gen}}{d\tau}=\frac{\Omega_{D-2}}{G_N L}P_{\infty}=\frac{\Omega_{D-2}}{G_N L}\sqrt{\tilde{U}(r_f)}\,,
\end{gather}
where $r=r_f$ is the radius of the local maximum. In certain parameter regimes, the effective potential may have more than one local maxima as shown in Fig.~\ref{fig:4d}.

\section{Codimension-one extremal slices in 4D}\label{sec:4d}
\subsection{Existence of local maxima}
\begin{figure}
    \centering
    \includegraphics[width=0.47\textwidth]{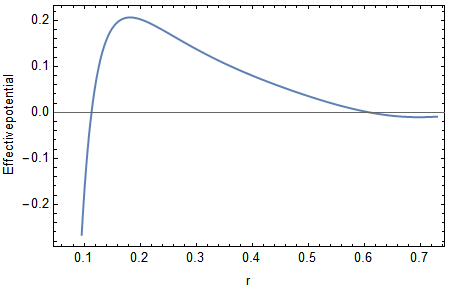}
    \includegraphics[width=0.48\textwidth]{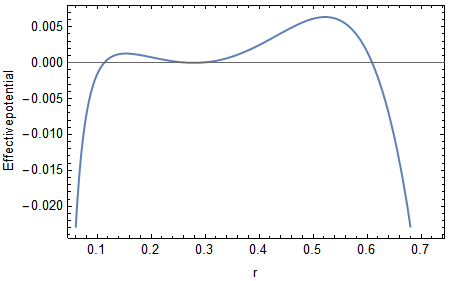}
    \includegraphics[width=0.47\textwidth]{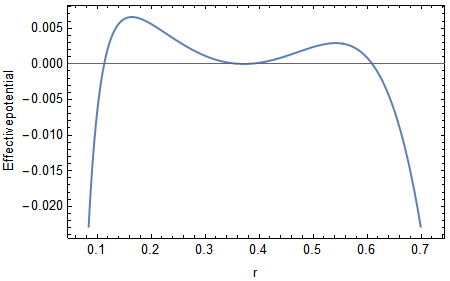} 
    \caption{Effective potential $\tilde U(r)$ as a function of radius $r$ in 4D. Left: $\lambda =-0.1$. Right: $\lambda=-0.01$. Bottom: $\lambda=-0.02$.  $\tilde U(r)$ vanishes on the inner and outer horizons and stays positive between them. This feature of the effective potential is universal in 4D Gauss-Bonnet black holes. For all, $L=w=1,\, \alpha=0.1.$}
    \label{fig:4d}
\end{figure}

For Gauss-Bonnet black hole, one simplest extension beyond the CV proposal in CAny is to add a higher curvature terms such as the square of Weyl tensors
\begin{gather}
    C^2=\frac{2}{D(D-1)}R^2-\frac{4}{D-1}R_{\mu\nu}R^{\mu\nu} + R_{\mu\nu\rho\sigma}R^{\mu\nu\rho\sigma}\,.
\end{gather}
The analysis is much simplified in 4D due to the different topology of Gauss-Bonnet-AdS black holes. Other choices of scalars include the generalization of CV conjecture in higher-curvature gravities by adding the extrinsic curvatures, which is investigated in detail in Ref.~\cite{Hernandez:2020nem}.  For spherically symmetric solutions of Gauss-Bonnet black holes in 4D, the square of Weyl tensor is given by
\begin{gather}
    C^2(r)=\frac{12 L^2 w^2 \left(L^2 \left(r^3+\alpha  w\right)-4 \alpha  r^3\right)^2}{\left(L^2 \left(r^4+4 \alpha  r w\right)-4 \alpha  r^4\right)^3}\,.
\end{gather}
One straightforward generalization of the CV proposal (which is recovered by setting $F_1=1$) is to pick $F_1(r)=1+\lambda L^4 C^2(r)$. This scalar function is the simplest nontrivial function in Einstein gravity in vacuum and is the most studied one. \footnote{The simpler choices of scalar functions such as $R^2$ and $R_{\mu\nu}R^{\mu\nu}$ do not contribute in the vacuum solutions of the Einstein gravity, therefore are not of particular interest in related studies.} Notably, the CAny proposals for Einstein gravity can be different for higher curvature gravities in general. One simplest extension is to lift the coupling constant $\lambda$ to a function of GB coupling $\alpha$. Furthermore, the CAny functional for higher curvature gravity may require extra extrinsic curvature terms as shown earlier for the CV proposal. The generalized holographic complexity is defined as the integration of the scalar function on the space volume that extremizes the integral, viz.,
\begin{gather}
    C_{gen}(\tau)=\frac{\Omega_{D-2}}{G_N L}\int_{\Sigma} d\sigma \sqrt{-f(r)\dot{v}^2+2\dot{v}\dot{r}}\, (1+\lambda(\alpha) L^4C^2(r)) \,.
\end{gather}
To ensure that the late time evolution of the generalized complexity increases linearly with the boundary time $\tau$, we plot the effective potential $\tilde U(r)$ as a function of radius $r$. For a reminder, the effective potential reads,
\begin{gather}
    \tilde{U}(r)=-f(r)F_1(r)^2\left(\frac{r}{L}\right)^{2(D-2)}=-f(r)\left(1+\lambda(\alpha) L^4 C^2(r)\right)^2\left(\frac{r}{L}\right)^{2(D-2)}\,.
\end{gather}
In principle, one can assume an equation of $\lambda(\alpha)$ and put constraints on its coefficients. To simplify our discussion, we drop the dependence of $\lambda(\alpha)$ on $\alpha$ and treat $\lambda$ as an arbitrary constant to be constrained in this study. The Penrose diagram of a Gauss-Bonnet-AdS black hole in 4D is similar to that of a charged black hole in that they are both equipped with two horizons. For spacetime regions inside the inner horizon or outside the outer horizon, i.e., $r<r_-$ or $r>r_+$, the lapse function is positive $f(r)>0$. For spacetime regions between the two horizons, $f(r)<0$. Therefore, the effective potential $\tilde U(r)$ is positive between the horizons and negative otherwise as shown in Fig.~\ref{fig:4d}. This ensures that the effective potential always has the local maximum $\tilde U(r_f)$ inside the horizon, which is the condition for the linear growth of complexity at late times. We will elaborate on this point further in the following discussion.

To see that any such observables in $D=4$ suffice as candidates for holographic complexity, we show that $\lim_{\tau\rightarrow \infty} P_\nu(\tau)=P_\infty$ is constant. Recall that the boundary time is
\begin{gather}
    \tau=-2 \int_{r_{\mathrm{min}}}^\infty dr \frac{P_\nu}{f(r)\sqrt{P_\nu^2-\tilde{U}(r)}}\,,
    \label{Eq:tau}
\end{gather}
where $r_{min}$ is determined by $P_\nu^2-\tilde{U}(r_{min})=0$. The lapse function $f(r)$ is continuous and nonzero between the two horizons. For $P^2_\nu<\tilde U(r_f)$ where $r_f$ is the radius of the local maximum of the effective potential, $\tilde U(r)\simeq P_\nu^2+\tilde U'(r_{min})(r-r_{min})$ in the vicinity of $r=r_{min}$. The integrand $\frac{P_\nu}{f(r)\sqrt{P_\nu^2-\tilde{U}(r)}}$ of Eq.~\eqref{Eq:tau} is regular everywhere except at $r=r_{min}$. Notice that for $U'(r_{min})\ne 0$, one can take the Cauchy principal value and the integral which gives the boundary time $\tau$ is finite. At late times, $\tau\rightarrow\infty$, and it corresponds to $P_\nu^2\rightarrow\tilde U(r_f)$. In this case, $\tilde U(r)\simeq P_\nu^2+\frac{1}{2}\tilde U''(r_{f})(r-r_{f})^2+O(r-r_f)^3$ near the local maximum $r=r_f$. One can see that as long as $r=r_f$ is the local maximum for $\tilde U(r)$, the function in the integrand is no longer integrable in the domain. In other words, the integral is divergent with the irregular point at $r=r_f$. This suggests that when $P_\nu$ approaches $\tilde U(r_f)$, the result of the integral which gives $\tau$ can take arbitrarily large values. Therefore, at late times $\tau \rightarrow \infty$, the conjugate momentum $P_\nu$ approaches a constant $P_\nu^2\rightarrow\tilde U(r_f)$, and the growth rate of the complexity is
\begin{gather}
    \lim_{\tau\rightarrow\infty} \frac{d C_{gen}}{d\tau}=\frac{\Omega_{D-2}}{G_N L}\sqrt{\tilde{U}(r_f)}\,.
\end{gather}
Therefore, CAny proposal has the natural realization at $D=4$ in Gauss-Bonnet-AdS black holes. \textit{No additional constraints or regularization for the scalar field is required for 4D Gauss-Bonnet gravity.}

\begin{figure}
    \centering
    \includegraphics[width=0.43\textwidth]{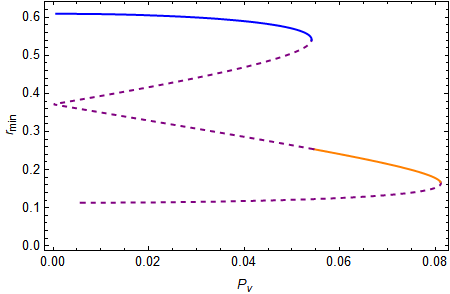} 
    \includegraphics[width=0.43\textwidth]{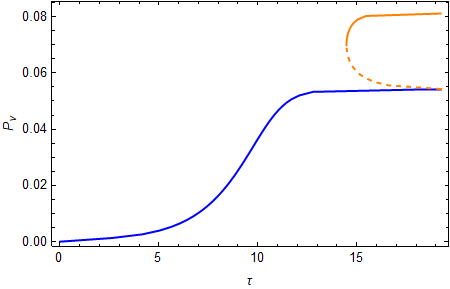}
    \caption{Conserved momentum when effective potential has two local maxima with global maximum locates inside. Left: the minimal radius $r_{\mathrm{min}}$ of the extremal surface vs conserved momentum at $\lambda=-0.02$. The dashed curves correspond to the hypersurfaces that do not have boundaries at $\Sigma_{\mathrm{CFT}}$. Right: boundary time vs conserved momentum at $\lambda=-0.02$. The dashed lines correspond to the hypersurface where  $P_\nu$ decreases and $r_{\mathrm{min}}$ increases as time evolves. The generalized volume evaluated on this slice is smaller than its counterpart measured along the blue curve. For both, $L=w=1,\, \alpha=0.1.$}
    \label{fig:2peaks}
\end{figure}

\subsection{Dipping and non-Dipping branches}
In the case of planar black holes in Einstein's theory, the effective potentials in diverge to infinity as $r\rightarrow 0$, and it is obvious that the equation $P_\nu^2=\tilde U$ always has a solution for arbitrarily large conjugate momentum $P_\nu$. It is not the case for 4D Gauss-Bonnet-AdS black holes where the effective potential approaches negative infinity as $r\rightarrow 0$. The three possible configurations of the effective potentials for a 4D Gauss-Bonnet-AdS black hole are shown in Fig.~\ref{fig:4d}. The ``dipping branch" in this scenario corresponds to the time evolution of $P_\nu$ approaching $P_\infty$ from above, i.e., $P_\nu$ decreases with the boundary time $\tau$ which is given by
\begin{gather}
    \tau = -2 \int^{r_{\mathrm{max}}}_{r_\mathrm{min}} dr \frac{P_\nu}{f(r)\sqrt{P_\nu^2-\tilde{U}(r)}}\,,
\end{gather}
where $r_{\mathrm{min}}$ the solution of $P_\nu=\tilde U(r)$. This process is represented by the orange solid lines in Fig.~\ref{fig:2peaks} (a) going from right to left and also by the dashed orange branch in Fig.~\ref{fig:2peaks} (b). One can compute numerically that the transitioning point from no dipping branch to the emergence of a dipping branch occurs at $\lambda\simeq-0.05$. It is worthwhile to point out that the extremal surfaces corresponding to the dipping branches in both the 4D Gauss-Bonnet gravity and the planar Einstein gravity do not yield the maximal holographic complexity at late times. This can be seen by comparing the late-time generalized volume evaluated on the dipping branch with that on the non-dipping branch
\begin{align}
    \lim_{\tau\rightarrow \infty} (C_{\mathrm{no-dip}}(\tau)-C_{\mathrm{dip}}(\tau))=\lim_{\tau\rightarrow \infty} \frac{\Omega_{D-2}}{G_N L} \int_{t=0}^{t=\tau} dt (P_\infty-\frac{\tilde U(r(t))}{P_\infty})\,.
\end{align}
At late times, as $P_\nu(\tau)\rightarrow P_\infty$ and the $\tilde U(r(t))\le P_\infty^2$, the right-hand side of the above equation remains positive. As in the case of planar black holes in Einstein gravity \cite{Belin:2022xmt}, the surface yielding the maximal complexity is always the one whose minimal radius approaches the local maximum of the effective potential. However, unlike the case of planar black holes in Einstein's theory, the extremal surface in the 4D Gauss-Bonnet-AdS black holes cannot probe the singularity since its boundary lies between the inner and outer horizons for arbitrarily large conjugate momenta. This is a reminiscence of a charged black hole \cite{Wang:2023eep}. The 4D Gauss-Bonnet black hole, unlike its counterparts in higher dimensions, has very similar causal structure as a RN-AdS black hole and this causal structure dictates qualitatively the motion of the extremal slice disregarding their differences.

\begin{figure}
    \centering
    \includegraphics[width=0.47\textwidth]{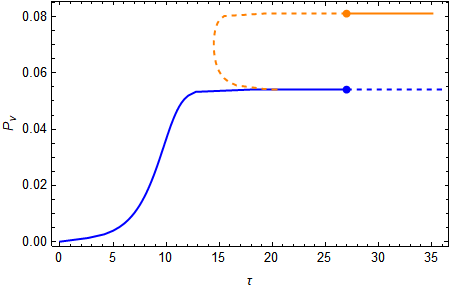} 
    \includegraphics[width=0.47\textwidth]{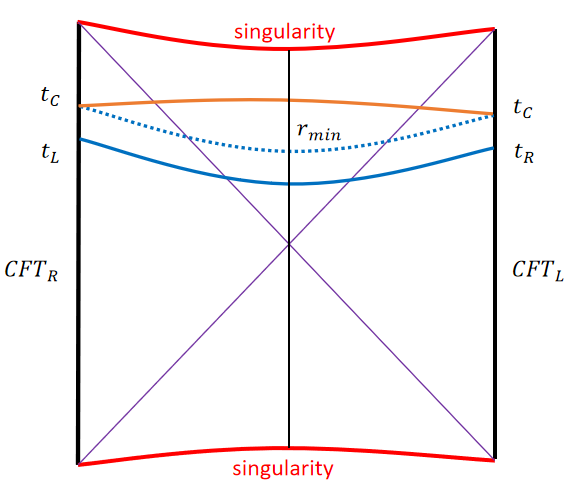} 
    \caption{Coexistence of two branches of extremal surfaces at $\lambda=-0.02$. Left: the boundary time $\tau$ vs the conserved momentum $P_\nu$. The two colors represent the two branches of the extremal surfaces. The blue curve represents the evolution from $(0,0)$. The generalized volume-complexity along the blue curve, which is proportional to the area below the curves, is dominating at the beginning of coexistence of two branches. After some critical time $t_C$, the volume-complexity computed from the orange curve dominates, and the conserved momentum jumps to a higher value discontinuous. Right: the Penrose diagram of the process. For both, $L=w=1,\, \alpha=0.1.$}
    \label{fig:jumps}
\end{figure}

\subsection{Phase transition of the extremal slice}
One feature of the holographic complexity in $4$D Gauss-Bonnet-AdS black holes is the appearance of multiple local maxima in the effective potential as illustrated in Fig.~\ref{fig:4d} (b) and (c). Consequently, the late-time behavior of the generalized volume is now characterized by two different branches, resulting in different rates of increase for the generalized volume-complexity. 

Notably, there is no continuous variation from one branch to the other along the trajectory of equations of motion. The time evolution of the conserved momentum $P_\nu$ with respect to the boundary time $\tau$ extends to an infinite future, ultimately approaching $P_{\infty,R}$, which is the asymptotic conserved momentum defined through Eq.~\eqref{Eq:tau} as $\tau\rightarrow \infty$. It is the value of $P_\nu$ at the right peak of the effective potential and is represented by the end value of the blue curve in Fig.~\ref{fig:jumps} (a). Along the trajectory of equations of motion, the evolution of the generalized volume-complexity is determined either by the left or by the right local maximum of the effective potential. Both trajectories of $P_\nu$ are smooth. 

However, according to the volume-complexity conjectures, the generalized volume-complexity $C_{gen}$ is determined by the maximum value obtained across all the branches. This requirement results in discontinuous jumps of the extremal surface from the blue curve to the orange one closer to the singularity at time $t_C$ as shown in Fig.~\ref{fig:jumps}. We remind that this ``phase transition" is not describable by the equations of motion of the extremal surfaces (Eq.~\ref{Eq:randv}) but rather is from the definition of the volume-complexity
\begin{gather}
     C_{gen}=\max\limits_\Sigma \left(\frac{\Omega_{D-2}}{G_N L}\int_\Sigma P_{\nu}(\tau)d\tau\right)\,,
\end{gather}
where $\Sigma$'s are the extremal surfaces. 

In Fig.~\ref{fig:jumps} (b), we demonstrate the two extremal surfaces in the Penrose diagram. At the critical time $t_C$, the volume-complexity evaluated on the two surfaces is identical. However, after the critical time, the surface closer to the singularity [the solid orange curve] exhibits a larger volume-complexity. As shown in Fig.~\ref{fig:2peaks} (a), the non-dipping branch, represented by the blue solid line on the top, has a lower generalized volume at $\tau\rightarrow\infty$. The extremal surface that generates the largest volume complexity jumps to the one ending at the left peak of Fig.~\ref{fig:4d}(c), and is represented by the orange solid line in Figs.~\ref{fig:2peaks} and \ref{fig:jumps}. This behavior resembles a phase transition and was briefly discussed in \cite{Wang:2023eep}. In particular, depending on the shape of the effective potential and in higher dimensional cases, this phase transition between different branches of $P_\nu-\tau$ diagrams can occur more than once. Overall, \textit{the late-time behavior is always dictated by the maximal peak of the effective potential $\tilde U(r)$ and the extremal surface after the transition always moves closer to the singularity}.

\section{Codimension-one extremal slices in 5D}\label{sec:5d}
In contrast to the four-dimensional case, Gauss-Bonnet black holes of dimensions higher than five exhibit the presence of only one horizon. For $d>4$, the analysis is similar and can be easily generalized, but it becomes much more complicated from Weyl tensor terms. Therefore, in this study, we will explicitly illustrate the five-dimensional case to demonstrate the constraints on the coupling parameter necessary for the existence of the extremal surface at late times.

\subsection{Existence proof of local maxima}
In 5D, the black hole has a minimal black hole mass condition, namely, $w>\alpha$. It possesses a single horizon, and its corresponding effective potential does not automatically admit a local maximum inside the horizon as observed in the 4D case. Notably, the lapse function $f(r)<0$ for the inside of the black hole $r<r_h$, which leads to the effective potential being positive inside the horizon, i.e., $\tilde U(r)>0$. Since $\tilde U(r)<0$ for $r>r_h$, it does not automatically ensures the existence of the local maximum as that in 4D. Therefore, the volume-complexity does not have the same simple universality as in 4D. 

\begin{figure}
    \centering
    \includegraphics[width=0.46\textwidth]{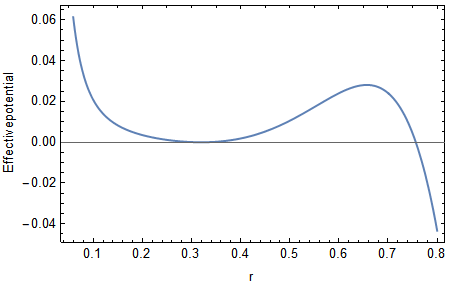}
    \includegraphics[width=0.46\textwidth]{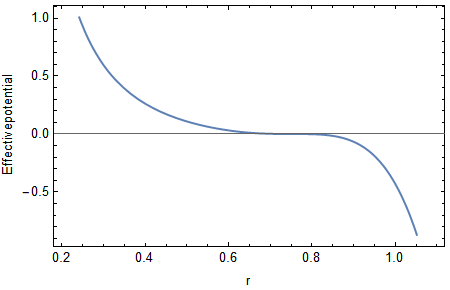}
    \includegraphics[width=0.46\textwidth]{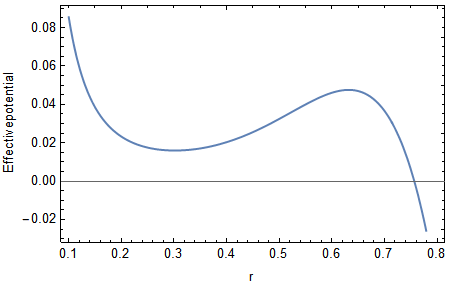}
    \includegraphics[width=0.46\textwidth]{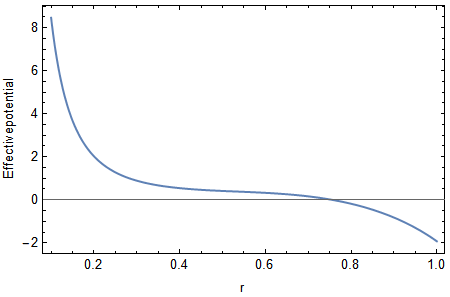}
    \caption{Effective potential $\tilde U(r)$ as a function of radius $r$ in 5D. (a) $\lambda=-0.0005>\lambda_c$. This corresponds to the existence of the local maximum inside the horizon. (b) $\lambda=-0.008877=\lambda_c$. This demonstrates the effective potential at exactly the critical point when the local maximum approaches the horizon and disappears. (c) $\lambda=0.001<\lambda_c'$. (d)  $\lambda=0.01>\lambda_c'$. For all, $L=w=1,\, \alpha=0.1.$ Unlike the case in $4$D, $\tilde U(r)\rightarrow \infty$ close to the singularity $r\rightarrow 0$.}
    \label{fig:5d}
\end{figure}

For the spherically symmetric solutions of Gauss-Bonnet black holes in 5D, the square of Weyl tensor is 
\begin{gather}
    C^2(r)= \frac{8 L^2 w^2 \left(L^2 \left(3 r^4+4 \alpha  w\right)-12 \alpha  r^4\right)^2}{r^4 \left(L^2 \left(r^4+4 \alpha  w\right)-4 \alpha  r^4\right)^3}\,.
\end{gather}
Therefore, the effective potential reads
\begin{gather}
    \tilde{U}(r)=-f(r)\left(1+\lambda(\alpha) \frac{8 L^6 w^2 \left(L^2 \left(3 r^4+4 \alpha  w\right)-12 \alpha  r^4\right)^2}{r^4 \left(L^2 \left(r^4+4 \alpha  w\right)-4 \alpha  r^4\right)^3}\right)^2\left(\frac{r}{L}\right)^{6}\,.
\end{gather}

In this case, it is difficult to directly solve for the local maximum of $\tilde U(r)$. Nevertheless, one may notice that the effective potential $\tilde U(r)\rightarrow\infty$ as $r\rightarrow 0$ as illustrated in Fig.~\ref{fig:5d}. The future null slice which hugs the singularity will always be the global maximum when evaluating the integral. This point will be delineated in detail later. 

For $\lambda(\alpha) <0$, we now show that the extremal surface always exists if the coupling parameter $\lambda(\alpha) $ lies within the range $\lambda_c<\lambda(\alpha) <0$ where $\lambda_c$ is the critical coupling to be determined later. For $\lambda(\alpha) <0$, $\lambda(\alpha)  C^2(r)\rightarrow -\infty$ near the singularity. Since $C^2(r)\rightarrow 0$ as $r\rightarrow \infty$, one can see that $\tilde U(r)/f(r)=0$ always has roots. For simplicity, we rewrite the square of Weyl tensor as
\begin{gather}
    C^2(\mathscr R)\propto \frac{(3\mathscr{R}+\beta)^2}{\mathscr R (\mathscr{R}+\beta)^3}\,,
    \label{eq:csquare}
\end{gather}
where $\mathscr R=(L^2-4\alpha)r^4$ and $\beta=4\alpha w L^2$. We remind that $(L^2-4\alpha)>0$ is required by the well-defined vacuum in the theory. One can check from the above formula Eq.~\eqref{eq:csquare} that $d C^2(\mathscr R)/d\mathscr R<0$ holds for all $\mathscr R>0$. Therefore, the equation for the effective potential 
\begin{gather}
    \tilde U(r)=-f(r)(1+\lambda(\alpha) C^2(r))^2(r/L)^{2(D-2)}=0
\end{gather}
has roots inside the horizon when the coupling parameter satisfies $\lambda_c<\lambda(\alpha) <0$, where $\lambda_c$ is the critical value of $\lambda$ that is determined by $\lambda_c C^2(r_h)=-1$ where $r_h$ is the radius of the horizon. The case $\lambda(\alpha) =\lambda_c$ corresponds to when the local minimal of the effective potential is zero at the horizon. Obviously, for $\lambda(\alpha) <\lambda_c$, the effective potential $\tilde U(r)>0$ for $r<r_h$ and it does not have a root inside the horizon. The effective potential $\tilde U(r)$ monotonically decreases inside the horizon, therefore, in this case it does not have a local maximum. In conclusion, the above analysis proves that when $\lambda_c<\lambda(\alpha)<0$, the requirement of linear growth for the observable $C_{gen}$ at late times is satisfied. In addition, the analysis also holds for Gauss-Bonnet black holes of higher-dimensions.

\begin{figure}
    \centering
    \includegraphics[width=0.5\textwidth]{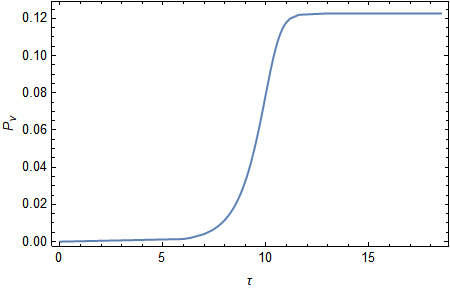}
    \caption{The change of conjugate momentum $P_\nu$ with boundary time $\tau$. $\lambda=-0.002$, $L=w=1,$ and $\alpha=0.1.$}
    \label{fig:pv-tau-5d1}
\end{figure}

For $\lambda(\alpha) \ge 0$, the effective potential satisfies $\tilde U(r)>0$ for $r<r_h$ so that it does not have a root inside the horizon. We notice that the effective potential $\tilde U(r)\propto -f(r)r^6(1+\lambda C^2(r))^2$. For $r^2 \ll \alpha$ (near the singularity), one has
\begin{align}
    \frac{d\left(-f(r) r^6\right)}{dr^4}&=\frac{-2 r^4 \left(\sqrt{-\frac{4 \alpha }{L^2}+\frac{4 \alpha  w}{r^4}+1}-1\right)-3 \alpha  \sqrt{r^4} \sqrt{-\frac{4 \alpha }{L^2}+\frac{4 \alpha  w}{r^4}+1}+6 \alpha  w -\frac{8 \alpha  r^4}{L^2}}{2 \alpha  \sqrt{-\frac{4 \alpha }{L^2}+\frac{4 \alpha  w}{r^4}+1}} \nonumber \\
    &\simeq 
    \frac{3\alpha  w- 3 \alpha \sqrt{\alpha w}}{\alpha \sqrt{4 \alpha  w}} r^2>0 \,.
\end{align}
Both the effective potential and its derivative with respect to the radius $r$ are positive inside the horizon and $\tilde U(r_h)=0$ on the horizon. This guarantees that when $\lambda(\alpha)=0$ there exists a local maximum $\tilde U(r)$ inside the horizon. Since the square of Weyl tensor $C^2(r)$ is a monotonically decaying function, it follows that there exists a maximal value of $\lambda_c'\ge 0$ such that: \\
$(1)$ for $0 \le \lambda(\alpha)  < \lambda_c'$, the effective $\tilde U(r)$ has a local maximum inside the horizon; \\$(2)$ for $\lambda(\alpha) >\lambda'_c$, $d\tilde U(r)/dr<0$ inside the horizon.\\
This completes the existence proof for the local maxima of the effective potential within a finite range of the coupling parameter. The maximal value $\lambda'_c$ depends non-trivially on $w\,,\alpha\,,L$ and can be determined numerically. The graphic illustration of the above results is shown in Fig.~\ref{fig:5d}. Combining the analysis above, we conclude that \textit{$C_{gen}$ is a well-behaved observable for the holographic complexity as long as $\lambda_c<\lambda(\alpha) <\lambda'_c$, where $\lambda_c=-1/C^2(r_h)<0$ and $\lambda'_c>0$.}

As shown in Fig.~\ref{fig:pv-tau-5d1}, when the coupling parameter is within the above-mentioned range, the conserved momentum monotonically increases with time and approaches a finite asymptotic value $P_\infty$. In this case, the generalized volume-complexity has a constant rate of increase at late times as $\lim\limits_{\tau\rightarrow\infty} \frac{d C_{gen}}{d\tau}=\frac{\Omega_{D-2}}{G_N L}P_\infty$.

\subsection{Couplings outside the allowed range}
In the previous section, we demonstrated that the desired behaviors of the {\vc} observable only manifest within a finite range of coupling parameters. This stands in contrast to the 4D case where no such constraints are necessary. However, even for the coupling parameters outside the desired range, we can just as well explore the behaviors of the conserved momentum and the extremal surfaces.

\begin{figure}
    \centering
    \includegraphics[width=0.47\textwidth]{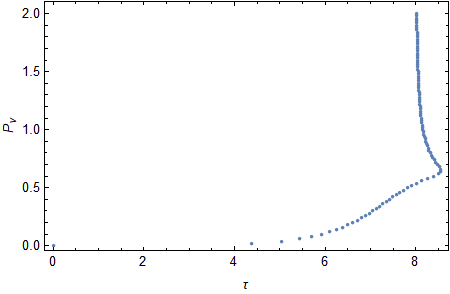}
    \includegraphics[width=0.47\textwidth]{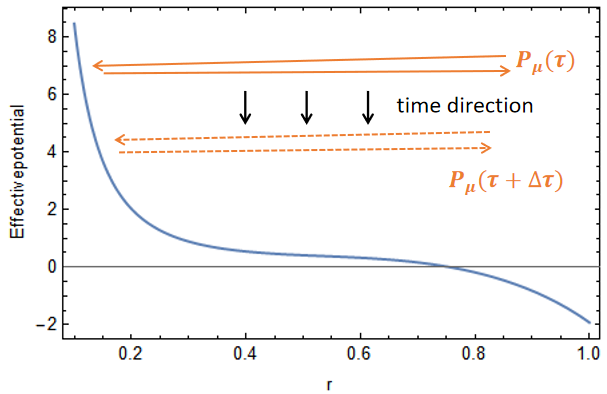}
    \includegraphics[width=0.47\textwidth]{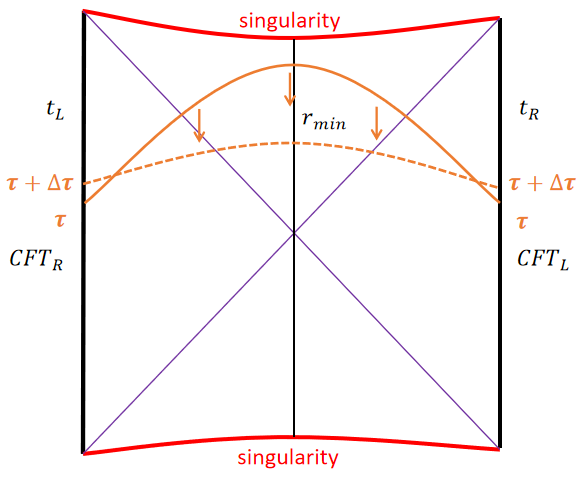}
    \caption{(a) Time evolution of the conserved momentum with $\lambda=0.01$, $L=w=1$, and $\alpha=0.1.$. (b) Diagrammatic illustration of the evolution of the conserved momentum at large values in (a). (c) The Penrose diagram for the process in (b) with $L=w=1,\, \alpha=0.1$ and $\lambda=-0.01$.}
    \label{fig:hugsing}
\end{figure}

First of all, we examine the evolution of the conserved momentum using Eq.~\eqref{Eq:tau}. For instance, consider a coupling parameter like $\lambda(\alpha)=0.01$, which exceeds the critical value $\lambda_c'$. As depicted in Fig.~\ref{fig:5d} (d), in this scenario, the {\ep} monotonically decreases. We evolve the extremal surface from $\tau=0$ and $P_\nu=0$. As shown in Fig.~\ref{fig:hugsing} (a), the evolution of the conserved momentum lasts only for a finite time before the extremal surface ceases to exist. On the other hand, if we start with an extremely large $P_\nu$, which corresponds to the extremal surface extending arbitrarily close to the singularity, $P_\nu$ rapidly decays to its final value and thereafter, no solutions exist any longer. This case is diagrammatically represented in Fig.~\ref{fig:hugsing} (b). In Fig.~\ref{fig:hugsing} (c), we depict the corresponding changes of the extremal surface in the Penrose diagram. Beginning with the extremal surface extending close to the singularity, the slice rapidly deforms into the one that is further away from the singularity before ceasing o exist. In this case, the generalized volume does not have a dual quantum complexity at late times.

\section{Discussion}
In the above analysis, the correction to the boundary time due to the presence of Gauss-Bonnet couplings is not considered \cite{An:2018dbz}. In general, for a more physical choice of boundary time that has the speed of light equal to one, the time coordinate should be shifted according to $t\rightarrow \frac{1}{N}t$ where $N=\sqrt{\frac{1}{2}\left(1+\sqrt{1-4\alpha/L^2}\right)}$. Therefore, the rate of change in complexity $\frac{dC}{dt}$ is modified by a constant multiplicative factor $N$. The Gauss-Bonnet correction to the boundary time decreases the complexity rate by a constant factor, but it does not affect our discussion. 

In the Complexity=Anything conjecture, a much wider class of observables is considered as viable candidates for the dual of holographic complexity. The core of evaluating the generalized volume, which is the complexity dual, is to find the extremal slice. In this study, we explored the CAny conjecture in four- and five-dimensional Gauss-Bonnet gravity in anti-de Sitter space. Specifically, we demonstrated the universality of the existence of the complexity dual at late times in 4D. Notably, the smallest radius to which the extremal surface extends always approaches a constant value at late times. This guarantees the finite asymptotic value of the conserved momentum and thus a constant increasing rate of the {\vc} at late times. We observed that starting from the boundary time $\tau=0$, the conserved momentum grows monotonically and the minimal radius of the extremal surface decreases monotonically to their corresponding finite asymptotic values at the local maximum of the effective potential. In cases where multiple local maxima co-exist, depending on the shape of the {\ep}, phase transitions of the extremal surface can emerge if the local maximum closer to the singularity has a larger peak value than the outermost one. Otherwise, the extremal slice follows a similar pattern as in the single-peak potential. Furthermore, we demonstrated that dipping slices, on which $P_\nu$ decreases with time, do not yield the largest generalized volume and can be disregarded in the computation. We argued that this feature extends beyond the Gauss-Bonnet gravity and is solely determined by the shape of the {\ep} of a specific model. 

The Gauss-Bonnet gravity in 4D and higher dimensions are dramatically different; the 4D gravity has time-like singularities and higher-D models all exhibit space-like singularities. In the 5D case, we showed that the universality of extremal slices at late times observed in 4D is not present, and only models with certain finite range of coupling parameters can serve as viable duals to the complexity. However, for coupling parameters outside the range, one can nevertheless evaluate the generalized volume according to the proposed formula. However, we found that the extremal surface only exists for a finite boundary time in this scenario and cannot replicate the constant growing feature of the complexity at late times. 

To conclude, we investigated the CAny proposal in the Gauss-Bonnet gravity and extended the discussion to include the phase transitions of the extremal slices as well as the deformations of the extremal slices in various scenarios in the Gauss-Bonnet-AdS gravity. Generalizations of CAny proposal in the Gauss-Bonnet as well as other higher curvature gravities can possibly include, for example, boundaries terms, extrinsic curvatures, the dependence of the coupling parameter on the Gauss-Bonnet and higher-curvature parameters, etc., which require further studies in the future. This may stimulate further quest for the understanding of the quantum nature of gravity and shed light to the interiors of the black holes. 

\acknowledgments
X.W would like to thank Yu-Sen An for helpful discussions. 
\bibliographystyle{JHEP}
\bibliography{zbib}

\end{document}